\documentclass[referee, usenatbib, useAMS]{biom}
\usepackage{graphicx}
\usepackage{bm, amsmath}
\usepackage{amsfonts}
\usepackage[T1]{fontenc}

\begin{document}

\title[Spatial Knockoff BVSR in GWAS]
{Spatial Knockoff Bayesian Variable Selection in Genome-Wide Association Studies}

\author{\\ Justin J. Van Ee$^{1,*}$\email{vanee002@colostate.edu},
Diana Gamba$^{2}$, Jesse R. Lasky$^{2}$, Megan  L. Vahsen$^{3}$, and \\ Mevin B. Hooten$^{4}$ \\
$^{1}$Department of Statistics, Colorado State University, Fort Collins, CO \\
$^{2}$Department of Biology, Pennsylvania State University, University Park, PA \\
$^{3}$Department of Wildland Resources and the Ecology Center, Utah State University, Logan, UT \\
$^{4}$Department of Statistics and Data Sciences, The University of Texas at Austin, Austin, TX}

\begin{abstract}
High-dimensional variable selection has emerged as one of the prevailing statistical challenges in the big data revolution. Many variable selection methods have been adapted for identifying single nucleotide polymorphisms (SNPs) linked to phenotypic variation in genome-wide association studies. We develop a Bayesian variable selection regression model for identifying SNPs linked to phenotypic variation. We modify our Bayesian variable selection regression models to control the false discovery rate of SNPs using a knockoff variable approach. We reduce spurious associations by regressing the phenotype of interest against a set of basis functions that account for the relatedness of individuals. Using a restricted regression approach, we simultaneously estimate the SNP-level effects while removing variation in the phenotype that can be explained by population structure. We also accommodate the spatial structure among causal SNPs by modeling their inclusion probabilities jointly with a reduced rank Gaussian process. In a simulation study, we demonstrate that our spatial Bayesian variable selection regression model controls the false discovery rate and increases power when the relevant SNPs are clustered. We conclude with an analysis of \textit{Arabidopsis thaliana} flowering time, a polygenic trait that is confounded with population structure, and find the discoveries of our method cluster near described flowering time genes.
\end{abstract}

 
\maketitle

\section{Introduction}

The availability of genetic data has proliferated over the past two decades because of increased medical research supported by governments and academic institutes as well as direct-to-consumer sequencing companies that have reduced the cost of collecting certain types of genetic data \citep{Lu2021}. In particular, advances in genotyping arrays have enabled the identification of millions of genome wide single nucleotide polymorphisms (SNPs) across thousands of individuals for relatively low costs \citep{Lu2021}. Genome-wide association studies (GWAS) leverage SNPs to investigate the genetic architecture of traits and provide insights into genotype-phenotype associations \citep{Visscher2012}. In the field of statistics, GWAS can be viewed as a high-dimensional variable selection problem where the phenotype of interest is the response and the SNPs are the covariates \citep{Guan2011}. Challenges to GWAS include controlling the number of SNPs that are falsely determined to be associated with phenotype (i.e., false discoveries) and efficiently obtaining posterior inference for the large parameter space. 

The vast majority of existing GWAS method rely on marginal testing where each SNP is tested independently for its association with the phenotype. A Benjamini–Hochberg procedure is then applied to the marginal test statistics to control the number of false discoveries \citep{Benjamini1995}. Because of the relatedness of individuals in the sample of genotypes, causal SNPs are often confounded with neutral genetic variation that does not influence phenotype but mimics the genetic architecture of the causal SNPs. To attenuate identification of these spurious associations, contemporary GWAS methods also include genotype random effects to account for population structure in a mixed model \citep{Kang2010, Zhou2012, Sul2018}. 

While single SNP testing methods have been instrumental for discovering thousands of genotype-phenotype associations \citep{Donnelly2008}, they have limitations. Many traits are polygenic and likely associated with large collections of SNPs. The individual effect of each SNP within collections is generally small, and marginal testing approaches can overlook collections of small effects \citep{Visscher2012b}. Likewise, because SNPs are tested independently, marginal testing cannot capture an epistatic genetic architecture \citep{Li2011, Lu2015}. Marginal testing also ignores the correlation structure among SNPs and thus overestimates the total proportion of variance explained by the selected SNPs \citep{Fridley2011, Guan2011}. 

A number of GWAS methods for simultaneously analyzing multiple SNPs have been proposed \citep{Guan2011, Li2011, Lu2015, Sesia2019, Sesia2020, Gu2021, Sesia2021}. Multiple SNP approaches use regularization to encourage sparsity and these include ridge and bridge regression, least absolute shrinkage and selection operator (LASSO), elastic net, and Bayesian variable selection regression (BVSR). Another commonality among multiple SNP GWAS approaches is a preprocessing of the response variable to reduce dependence because of population structure, which can decrease the number of false discoveries and improve power \citep{Price2006}. Nonetheless, in finite samples, preprocessing does not provide theoretical controls for the false discovery rate (FDR). 

An advantage of the BVSR approach is the flexibility to incorporate prior information into the selection of variables that reflect their latent structure. Such models have gained particular attention in biology where the inclusion indicators of covariates have been modeled as Markov random fields \citep{Li2010, Vannucci2010}. \cite{Stingo2011} developed a BVSR model that incorporated the relationship between genes and their membership to biochemical pathways to improve understanding of their expression levels on a phenotype of interest. Several studies have modeled voxel selection indicators jointly with an Ising prior to control the sparsity and smoothness of selection in functional MRI studies of the brain \citep{Li2010}. In both contexts, modeling the selection indicators jointly with a Markov random field reduced the false positive rate and improved predictive performance relative to models that did not account for latent spatial structure among the predictors. 

We develop a BVSR model that uses knockoff variables \citep{Barber2015} to achieve FDR control in finite samples. \cite{Candes2018} first proposed the knockoff variables in the context of BVSR, but extensions and applications to real datasets are currently sparse \citep{Gu2021}. \cite{Guan2011} developed a computationally efficient BVSR model for variable selection in GWAS that produced better power and predictive performance compared with standard single-SNP analyses and LASSO regression. Our contribution is an extension to the BVSR model proposed by \cite{Guan2011} with three notable additions. First, we implement a restricted regression framework to account for population structure in a cohesive linear model for environmental, relatedness, and SNP effects. Second, we modify existing BVSR approaches to control the FDR using knockoff variables. Third, we improve power to detect relevant variants by modeling SNP inclusion probabilities jointly with a reduced rank Gaussian process. While each of these individual extensions has received considerable attention in the existing literature, our work represents the first approach to simultaneously implement them in a single cohesive framework. 

To assess the performance of our spatial BVSR model in a realistic settling, we analyzed the genetic factors influencing flowering time in $1058$ wild accessions of \textit{Arabidopsis thaliana}. Because of range expansions, bottlenecks, and multiple reintroductions to its invaded ranges, wild populations of \textit{Arabidopsis thaliana} have complex population structure with high levels of admixture \citep{Shirsekar2021} that make GWAS challenging. We focused on flowering time for our analysis because the trait has a polygenic architecture \citep{Zan2018} with $282$ described genes \citep{Brachi2010}. Flowering time plays a central role in ecological adaptation and has been extensively studied for \textit{Arabidopsis thaliana} \citep{Shindo2005, Atwell2010, Brachi2010, Brachi2013, Alonso2016, Zan2018}. Adaptive traits under intensive selective pressure are generally polygenic \citep{Campbell2021} as evidenced by the hundreds of genes involved in regulating flowering time in \textit{Arabidopsis thaliana} \citep{Brachi2010}. Traditional single SNP analyses of \textit{Arabidopsis thaliana} flowering time have shown limited success, only revealing a few significant associations \citep{Shindo2005, Atwell2010, Brachi2010, Brachi2013, Alonso2016}. We investigate whether our spatial knockoff BVSR model can overcome the shortcomings of traditional approaches.

\section{Methods}\label{Methods_Section}

\subsection{Bayesian Variable Selection Regression}

We developed a model for identifying SNP-phenotype associations in GWAS. Consider the generalized linear model (GLM) 
\begin{align}
    \bm{y}&\sim{}[\bm{y}|\bm{\mu},\phi] \label{BVSR_glm_mod}, \\
    f(\bm{\mu})&\sim\mathcal{N}(\bm{X}\bm{\beta}+\bm{Z}\bm{G}\bm{u}, \sigma^2_{e}\bm{I}), \label{BVSR_lin_mod}
\end{align}
where we use the bracket notation, $[\cdot]$, to denote probability distributions \citep{Gelfand1990},  $\bm{y}$ is a vector of phenotypes observed across $n$ individuals with mean $\bm{\mu}$, $\phi$ is a set of additional parameters, possibly empty, in the data model, $\bm{X}$ is a $n\times p$ matrix of covariates including an intercept, $\bm{G}$ is a $n_g\times n_u$ genotype matrix containing the number of alleles each genotype has at a particular locus for the $n_u$ SNPs in the sample of $n_g$ genotypes, and $\bm{Z}$ is a matrix of $1$s and $0$s that links the $n_g$ genotypes in the sample to the $n$ observed phenotypes. As parameterized, the GLM assumes an additive effect at each loci, but we could also test for dominant and recessive effects at each loci if we added another $n_u$ covariates \citep{Li2011}. 

We assume $n_u>>n$ and that each SNP $j\in\{1,\dots,n_u\}$ only explains a small proportion of the total variation in $\bm{y}$ \citep{Visscher2012b}. Furthermore, we assume that only a small fraction of the $n_u$ sequenced SNPs are associated with phenotype and will refer to these SNPs as ``relevant'' despite that these SNPs might not be causally related with phenotype. We seek to identify the relevant SNPs while controlling the number of SNPs we falsely conclude are important. 

We adopted a BVSR approach and specified spike-and-slab priors \citep{Mitchell1988} for the SNP effects,
\begin{align}
\label{spike-and-slab prior}
    u_j & \sim\begin{cases}
                            \mathcal{N}(0,\sigma^2_{a}), & \text{for $\nu_j=1$} \\
                            0, & \text{for $\nu_j=0$}
                         \end{cases}, \\
    \nu_j & \sim\text{Bernoulli}(\pi_j).
\end{align}
The indicator variables, $\nu_j$, act as switches for adding and removing SNPs from the model with $\nu_j=1$ indicating that SNP $j$ improves predictive performance for phenotype $\bm{y}$. The posterior means of the $\nu_j$ are called posterior inclusion probabilities and help assess whether a SNP is associated with variation in phenotype. In the context of GWAS, a BVSR approach is appealing because many SNPs are assumed to be unassociated with phenotype and setting $u_j=0$ for these SNPs is theoretically justifiable \citep{Fridley2009}. The spike-and-slab prior removes unimportant SNPs from the model, which both reduces runtime \citep{Lu2015} and improves posterior inference for the causal SNP effects \citep{Guan2011}.

\cite{Guan2011} introduced a flexible prior for the variance of the SNP-level effects, $\sigma^2_{a}$, that applies greater shrinkage for more complex models with many non-zero SNP effects. We let $s_j^2$ represent the sample variance of SNP $j$ in allelic state and $h\sim\mathcal{U}(0,1)$. \cite{Guan2011} induced a conditional prior distribution for $\sigma^2_{a}$, $[\sigma^2_{a}|\bm{\nu}, \bm{s}^2, h]$, by defining 
\begin{align}
    \sigma^2_a(\bm{\nu}, \bm{s}^2, h)=\frac{h}{1-h}\frac{1}{\sum_{j:\nu_j=1}s_j^2}.
\end{align}
Regardless of how many SNPs are estimated as non-zero, the adaptive prior holds the proportion of variance explained by all SNPs constant \citep{Guan2011}. The prior is also heavy tailed, an attractive property for minimally informative variance parameters in hierarchical models \citep{Gelman2006}, with density proportional to $f(h)=\frac{1}{(1+h)^2}$. We adopted the prior specification of \cite{Guan2011} for the variance of the non-zero SNP effects, $\sigma^2_{a}$.

\subsection{Accounting for Population Structure}

An implicit assumption of equation (\ref{BVSR_lin_mod}) is that all $n_g$ genotyped individuals are unrelated. When the genotyped individuals are related, the observed phenotypes are no longer independent, and the BVSR model may select many SNPs that are spuriously associated with $\bm{y}$ \citep{Sul2018}. In single SNP methods, individual SNP-level effects are estimated jointly with a set of $n_g$ genotype random effects that correct for population structure \citep{Kang2010, Zhou2012, Sul2018}. For the multiple SNP methods, a common approach is to first decorrelate the observations by regressing $\bm{y}$ against a set of basis functions that describe the relatedness among individuals \citep{Guan2011, Li2011, Lu2015, Sesia2019}. The residuals, $\bm{e}$, from the regression then replace $\bm{y}$ in equation (\ref{BVSR_glm_mod}). 

A downside of the step-wise procedure used for multiple SNP methods is that it does not propagate the uncertainty associated with estimating the basis function coefficients into the estimates of the other model parameters. We simultaneously estimate the environmental, SNP, and relatedness effects by extending the original GLM, equation (\ref{BVSR_lin_mod}), to the following generalized linear mixed model (GLMM), 
\begin{align}
    f(\bm{\mu})&\sim\mathcal{N}(\bm{R}\bm{\theta}+\bm{X}\bm{\beta}+\bm{Z}\bm{G}\bm{u}, \sigma^2_{e}\bm{I}),\label{lin_mod_pop}
\end{align}
where $\bm{R}$ is a matrix of $n_R$ basis functions that provide a low-dimensional representation of the relatedness among individuals. Low-dimensional representations can be obtained from a singular value decomposition (SVD) of either the genotype or kinship matrix \citep{Price2006}. The choice of basis function type and number will depend on the relatedness of individuals in the sample.

The coefficients $\bm{\theta}$ and $\bm{u}$ are confounded in equation (\ref{lin_mod_pop}) because of overlap in the column spaces of $\bm{R}$ and $\bm{Z}\bm{G}$. Variation in the phenotype can be explained by both SNP-level effects as well as population structure. To alleviate confounding, we adopted a restricted regression approach \citep{Reich2006} and reparameterized the model as 
\begin{align}
    f(\bm{\mu})&\sim\mathcal{N}(\bm{R}\bm{\theta}+\bm{X}\bm{\beta}+\bm{K}\bm{u}, \sigma^2_{e}\bm{I}),\label{lin_mod_pop_res}
\end{align}
where $\bm{K}=(\bm{I}-\bm{P}_{\bm{R}})\bm{Z}\bm{G}$ and $\bm{P}_{\bm{R}}=\bm{R}(\bm{R}'\bm{R})^{-1}\bm{R}'$ is the projection matrix onto the column space of $\bm{R}$. The restricted model, equation (\ref{lin_mod_pop_res}), gives priority to the relatedness effects, $\bm{\theta}$, to explain all the contested sources of variation in the response \citep{Reich2006}. Conceptually, the restricted model is similar to the step-wise procedures in that the SNP-level effects are estimated from the residual variation in $\bm{y}$ left over from regressions that correct for population structure.  Because SNPs are measured for an effect on the phenotype that is not explained by population structure, both procedures lack power to detect the relevant variants that are confounded with the relatedness of individuals \citep{Klasen2016}. The benefit of the restricted model is that all parameters are estimated simultaneously such that their uncertainties are accounted for in the model. For observational data, the environmental and genetic effects may also be confounded, and restricted regression could be used for $\bm{\beta}$ (i.e., $\bm{K}=(\bm{I}-\bm{P}_{\bm{R}})(\bm{I}-\bm{P}_{\bm{X}})\bm{Z}\bm{G}$).

\subsection{Controlling False Discovery Rate Via Knockoffs}

\cite{Barber2015} introduced the knockoff filter for general variable selection and several extensions have been proposed in the context of GWAS \citep{Candes2018, Sesia2019, Sesia2020, Sesia2021}. Knockoffs are synthetic variables constructed such that they closely resemble the correlation structure of the original predictors but are independent of the response. The knockoff filter leverages the synthetic variables to calibrate the selection procedure such that the FDR is controlled at the desired level. The FDR control is exact even in finite sample settings regardless of the design or covariates, the number of variables in the model, or the noise to signal ratio \citep{Barber2015}. The knockoff filter can be applied in a wide range of models, but for brevity, we describe the method in the context of our BVSR model. 

Consider the augmented model, 
\begin{align}
    \bm{y}&\sim{}[\bm{y}|\bm{\mu},\phi], \\
    f(\bm{\mu})&\sim\mathcal{N}(\bm{R}\bm{\theta}+\bm{X}\bm{\beta}+\bm{K}\bm{u}+\tilde{\bm{K}}\tilde{\bm{u}}, \sigma^2_{e}\bm{I}),\label{knockofflinmod}
\end{align}
where $\tilde{\bm{K}}=(\bm{I}-\bm{P}_{\bm{R}})\bm{Z}\tilde{\bm{G}}$ and $\tilde{\bm{G}}$ is the knockoff copy of ${\bm{G}}$. We describe the properties of $\tilde{\bm{G}}$ in Section \ref{ModelXKnockoffs}. Following \cite{Candes2018}, we express a joint spike-and-slab model for the original and knockoff SNP effects, $(u_j, \tilde{u}_j)$, as 
\begin{align}
    (u_j, \tilde{u}_j) & \sim (\delta_j, \tilde{\delta}_j)\mathcal{N}(0,\sigma^2_{a}), \\
    (\delta_j, \tilde{\delta}_j) & \sim \begin{cases}
    \text{Categorical}(0.5,0.5), & \text{for $\nu_j=1$} \\
    (0,0), & \text{for $\nu_j=0$} 
    \end{cases}, \\
    \nu_j & \sim\text{Bernoulli}(\pi_j).
\end{align}
We define the quantity $w_j=\delta_j-\tilde{\delta}_j$ and denote its posterior mean $\bar{w}_j=\mathbb{E}[w_j|\bm{y}]$. True discoveries are indicated by $\bar{w}_j>0$ whereas false discoveries correspond to $\bar{w}_j\le0$. We also redefine the variance of the original and knockoff SNP effects
\begin{align}
    \sigma^2_a(h, \bm{\delta}, \tilde{\bm{\delta}}, \bm{s}^2, \tilde{\bm{s}}^{2})&=\frac{h}{1-h}\frac{1}{\sum_{j:\delta_j=1}s_j^2+\sum_{j:\tilde{\delta}_j=1}\tilde{s}_j^2},
\end{align}
so that $\sigma^2_a$ shrinks as either type of variable is added to the model. 

Suppose we select all SNPs having $\bar{W}_j>t^{\star}$ for some $t^{\star}\in(0,1)$. We let $\hat{S}\subset\{1,\dots,n_u\}$ be the subset of SNPs selected. The FDR of this procedure is 
\begin{align}
    \text{FDR}=\mathbb{E}\left[\frac{\#\{j:\text{$u_j=0$ and $j\in\hat{S}$}\}}{\#\{j:\text{$j\in\hat{S}$}\}\vee1}\right],
\end{align}
where we use the notation $a\vee b$ to denote $\max\{a,b\}$. The goal is to chose $t^{\star}$ as small as possible subject to the constraint that FDR is controlled below some prespecified threshold $q$. \cite{Barber2015} proved the optimal $t^{\star}$ is given by 
\begin{align}\label{knockoffthreshold}
    t^{\star}=\text{min}\left\{t\in(0,1):\frac{1+\#\{j:W_j\le -t\}}{\#\{j:W_j\ge t\}\vee1}\le q\right\}.
\end{align}
The threshold, $t^{\star}$, in equation (\ref{knockoffthreshold}) controls the expected number of false discoveries, but note that for any one analysis the observed proportion of false discoveries may exceed q. 

\subsubsection{`Model-X' Knockoffs}\label{ModelXKnockoffs}

Knockoff variables as proposed in \cite{Barber2015} are constructed geometrically and only valid if $n_u<2n_g$. \cite{Candes2018} introduced probabilistically constructed `Model-X' knockoffs for high-dimensional variable selection. Given a family of random variables $\bm{g}=(g_1,\dots,g_{n_u})'$, a `Model-X' knockoff, $\tilde{\bm{g}}=(\tilde{g}_1,\dots,\tilde{g}_{n_u})'$, satisfies two properties: 
\begin{enumerate}
    \item for any subset $S\subset\{1,\dots,n_u\}$, $\left(\bm{g},\tilde{\bm{g}}\right)_{\text{swap}(S)}\stackrel{d}{=}\left(\bm{g},\tilde{\bm{g}}\right)$,
    \item $\tilde{\bm{g}}\perp \!\!\! \perp\bm{y} | \bm{g}$,
\end{enumerate}
where $\stackrel{d}{=}$ denotes equality in distribution and $\left(\bm{g},\tilde{\bm{g}}\right)_{\text{swap}(S)}$ is obtained by swapping the variables $g_j$ and $\tilde{g}_j$ for all $j\in S$. 

A trivial knockoff satisfying criteria 1 and 2 is $\tilde{\bm{g}}=\bm{g}$. This knockoff would be of little practical use because $\bar{w}_j\approx0$ for all $j$ yielding no power. As a more relevant example, if the variables follow a Gaussian distribution, $\bm{g}\sim\mathcal{N}(\bm{0}, \bm{\Sigma})$, one possible knockoff construction is 
\begin{align}
    (\bm{g},\tilde{\bm{g}})\sim\mathcal{N}(\bm{0}, \bm{H})\text{, where } \bm{H}=\begin{pmatrix}
\bm{\Sigma} & \bm{\Sigma}-\text{diag}(\bm{s}) \\
\bm{\Sigma}-\text{diag}(\bm{s}) & \bm{\Sigma} 
\end{pmatrix},
\end{align}
and $\text{diag}(\bm{s})$ is any diagonal matrix selected in such a way that the joint covariance matrix is positive definite. In general, knockoffs become more powerful as the absolute pairwise correlation between each variable and its knockoff copy decreases. 

\subsection{Spatially Structured Inclusion Probabilities}\label{spatialinclusion}

Traditional approaches for BVSR have treated the inclusion probabilities as a fixed hyperparameter, $\pi_j=\pi$ for $j=1,\dots,n_u$ \citep{Mitchell1988, George1993, Smith1996, Raftery1997, Brown2002}. While fixing $\pi_j$ may be appropriate for some analyses, in many cases, the sparsity may not be known even to the correct order of magnitude.  \cite{Guan2011} specified the log uniform prior 
\begin{align}\label{log_uniform_prior}
    \log(\pi)\sim\mathcal{U}(a,b),
\end{align}
with $a=\log(1/n_u)$ and $b=\log(M/n_u)$, so that the lower and upper limits on $\pi$ correspond to an expectation of $1$ and $M$ variables included in the model, respectively, where $M$ is a hyperparameter. The log uniform prior puts approximately equal probability on different magnitudes of sparsity (e.g., $10^{-3}, 10^{-4},$ and $10^{-5}$) whereas a uniform prior would favor larger magnitudes. In GWAS, prior information on the sparsity of relevant SNPs is rarely available making the log uniform prior an appealing choice. 

\begin{figure}[ht]
\centering
\includegraphics[scale=0.60]{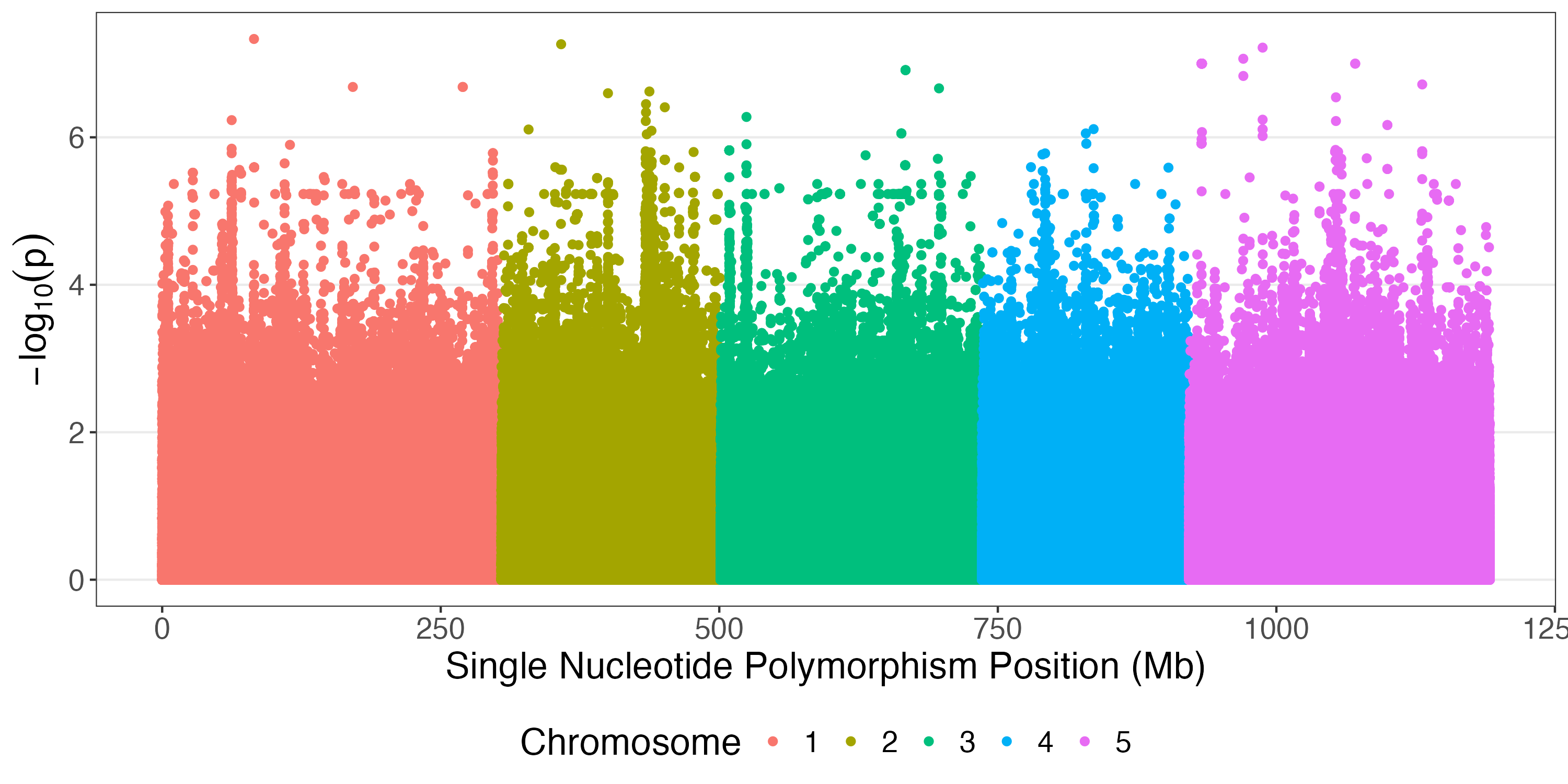} 
\caption{Manhattan plot of marginal associations with flowering time for $\approx 7$ million single nucleotide polymorphisms (SNPs) in \textit{Arabidopsis thaliana}. The vertical axis is the negative log (base 10) of the p-values from t-tests for univariate linear regressions of each SNP with flowering time. No marginal associations were significant at a false probability level of $0.05$ with Bonferroni correction.} 
\label{Manhattan_Arabidopsis}
\end{figure}

\cite{Guan2011} showed the log uniform prior provides accurate posterior inference for a wide range of sparsities, but assuming a common sparsity across all chromosomes and within each chromosome may not always be appropriate. For many traits, the SNPs associated with phenotype can be restricted to relatively few clustered loci. In humans, for example, roughly $95\%$ of disease-causing mutations occur in exonic regions that only make up $1-2\%$ of the entire genome \citep{Posey2019}. If a trait is monogenic, all causal SNPs could fall within a gene of less than 20 Kilobases (Kb). Figure \ref{Manhattan_Arabidopsis} shows a Manhattan plot of SNP marginal associations with flowering time in \textit{Arabidopsis thaliana}. Flowering time in \textit{Arabidopsis thaliana} is known to have a complex polygenic architecture \citep{Zan2018}, and this is reflected by many moderate signals expressed throughout the genome. We also see spatial structure among signals with the largest marginal associations often occurring in clusters. 

Motivated by the clustering of causal SNPs in GWAS, we developed a reduced rank approach for accommodating spatial structure. Within each chromosome, we modeled the dependence in the SNP inclusion probabilities using the conditional autoregressive (CAR) process, 
\begin{align}\label{logit_normal_prior}
    \text{logit}(\bm{\pi})\sim\mathcal{N}(\mu_{\pi}\bm{1}, \tau\bm{L}(\rho)),
\end{align}
where $\bm{L}(\rho)=\left(\text{diag}(\bm{A}\bm{1})-\rho\bm{A}\right)^{-1}$, $\bm{A}$ is a proximity matrix \citep{Hooten2019}, and we set $\rho\rightarrow1$ to induce an approximate intrinsic CAR (ICAR) process \citep{VerHoef2018}. For the proximity matrix $\bm{A}$, we specified the neighborhood structure,
\begin{align}
    a_{ij}=\begin{cases} 
  0, & \text{$i=j$} \\  
  1, & \text{$i\ne j$, $d_{ij}\le d_{\text{thresh}}$} \\  
  0, & \text{$i\ne j$, $d_{ij}\ge d_{\text{thresh}}$} \\  
                    \end{cases},
\end{align}
where $d_{ij}$ is the distance between SNP $i$ and $j$ in bases, and $d_{\text{thresh}}$ is selected \textit{a priori}. 

Inverting $\bm{L}(\rho)$ is prohibitive for the large $n_u$ typically encountered in GWAS. To reduce computational burden, we used a basis function approach for incorporating spatial dependence \citep{Hefley2017basis}, and let 
\begin{align}
    \text{logit}(\bm{\pi}) & =\mu_{\pi}\bm{1}+\bm{B}\bm{\alpha}, \\
    \bm{\alpha} & \sim\mathcal{N}(\bm{0}, \tau\bm{B}'\bm{L}(\rho)\bm{B}), \label{alphaequation}
\end{align}
where $\bm{B}$ is a basis expansion of $\bm{L}(\rho)$. We adopted the basis expansion proposed by \cite{Hughes2013} and let $\bm{B}=\bm{Q}\bm{\Lambda}$, where $\bm{Q}\bm{\Lambda}\bm{Q}'$ is the spectral decomposition of the Moran operator matrix $\bm{M}$ of $\bm{A}$,
\begin{align}
    \bm{M} & =\frac{n_u(\bm{I}-\frac{1}{n_u}\bm{1}\bm{1}')\bm{A}(\bm{I}-\frac{1}{n_u}\bm{1}\bm{1}')}{\bm{1}'\bm{A}\bm{1}}.
\end{align}
The construction of $\bm{B}$ reduces confounding with $\mu_\pi$ and captures spatial dependence in the arrangement of SNPs characterized by $\bm{A}$. We reduced computational burden by using the first $n_{\alpha}$ vectors from the basis expansion. The accuracy of the approximation decreases with fewer vectors included but selecting $n_{\alpha}<<n_u$ generally has negligible effects on posterior inference \citep{Hughes2013}. 

Nearby SNPs can have similar associations because of two distinct biological mechanisms. Protein translation leads SNPs located within a gene to be causally linked to a phenotype of interest \citep{Wang2010}. Nearby SNPs also tend to have similar marginal associations as a result of strong mutual correlations resulting from limited recombination during meiosis, a pattern known as linkage disequilibrium \citep{Jorde2000}. Because the vast majority of eukaryote DNA is non-exonic \citep{VanStraalen2011}, the effects of linkage disequilibrium on SNP clustering likely swamp those of protein translation such that most clusters will be a mix of causally and spuriously associated SNPs. We address this issue by first pruning the genotype matrix and selecting representatives from collections of SNPs in linkage disequilibrium (i.e., SNPs having a mutual correlation above some prespecified threshold). The details of our variable pruning procedure are given in Web Supplement C. Having attenuated the effects of linkage disequilibrium with variable pruning, the spatially structured inclusion probabilities can improve power to detect variants with weak marginal associations that are located near other relevant SNPs \citep{Benjamini2007, Wang2010}. 

\subsection{Implementation}\label{Implementation_Section}

Two challenges in implementing the described models were efficiently obtaining posterior inference for the high-dimensional parameter space and generating the Model-X knockoffs. We obtained a posterior sample for all quantities using Markov chain Monte Carlo (MCMC). The crux of making MCMC computationally feasible was to avoid sampling the parameters with dimension $n_u$ each MCMC iteration. In Web Supplement A, we describe several sampling strategies we used to improve computation. Constructing knockoffs in the context of GWAS studies is difficult because $\bm{G}$ in equation (\ref{knockofflinmod}) is both high-dimensional and also non-Gaussian. \cite{Sesia2019} developed an algorithm for generating knockoff copies of SNPs using a hidden Markov model (HMM). In the framework of \cite{Sesia2019}, point estimates of the HMM parameters are first estimated with an expectation–maximization algorithm in \texttt{fastPHASE} \citep{Scheet2006}, and then used for sampling cheap knockoff copies within the HMM framework. Because the underlying distribution of the SNPs is unknown in practice, the generated knockoffs are approximate. \cite{Sesia2019} showed that for both simulated and real datasets the HMM knockoffs control FDR while maintaining high power, and we used the algorithm of \cite{Sesia2019} for generating our Model-X knockoffs. 

\section{Results}\label{Results_Section}

We fit a non-spatial BVSR model with log uniform prior on $\pi$ as in equation (\ref{log_uniform_prior}) and our spatial BVSR model to a variety of simulated datasets and flowering time observations for $1058$ wild accessions of \textit{Arabidopsis thaliana}. We adopted a normal data model (i.e. $f(\bm{u})=\bm{u}$ in equation (\ref{BVSR_lin_mod})) for all analyses. Across simulations we varied the noise to signal ratio, degree of spatial structure, and the inclusion of both population structure and linkage disequilibrium. We provide full model statements, data simulation and preprocessing details, and model fitting statistics for the simulated and real datasets in Web Supplements B and C, respectively. 

\begin{figure}[ht]
\centering
\includegraphics[scale=0.8]{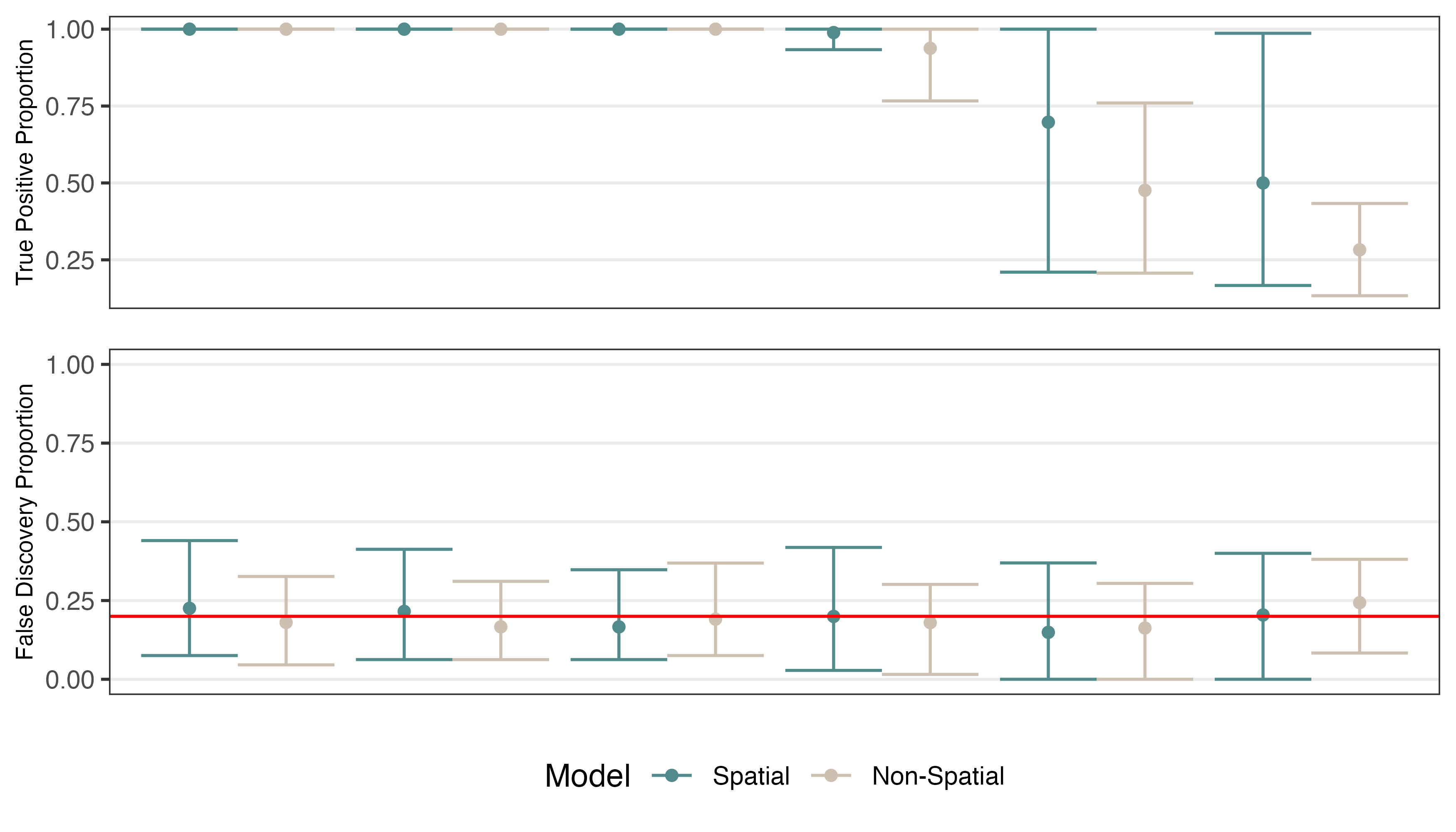} 
\caption{Means, $5$th, and $95$th quantiles of false discovery and true positive proportions. The true positive proportion is the number of causal SNPs correctly identified divided by $30$, the true number of simulated relevant SNPs. We varied the noise to signal ratio by setting $\sigma^2_e=1$ in equation (\ref{BVSR_lin_mod}) and letting $u_{j}=1,\dots,6$ in equation (\ref{spike-and-slab prior}) for the $30$ relevant SNPs. Results are pooled across six clustering regimes: all relevant SNPs in $1$, $3$, $5$, $10$, $15$, and $30$ clusters, respectively. We simulated $10$ datasets for each clustering regime and noise to signal ratio, for a total of $360$ datasets. All simulated datasets are for $n_u=20,000$ uncorrelated SNPs (no linkage disequilibrium) observed across $n=1,000$ unrelated (no population structure) individuals. The red horizontal line depicts the targeted false discovery proportion of $q=0.20$.} 
\label{errorbar_SNP}
\end{figure}

\begin{figure}[ht]
\centering
\includegraphics[scale=0.8]{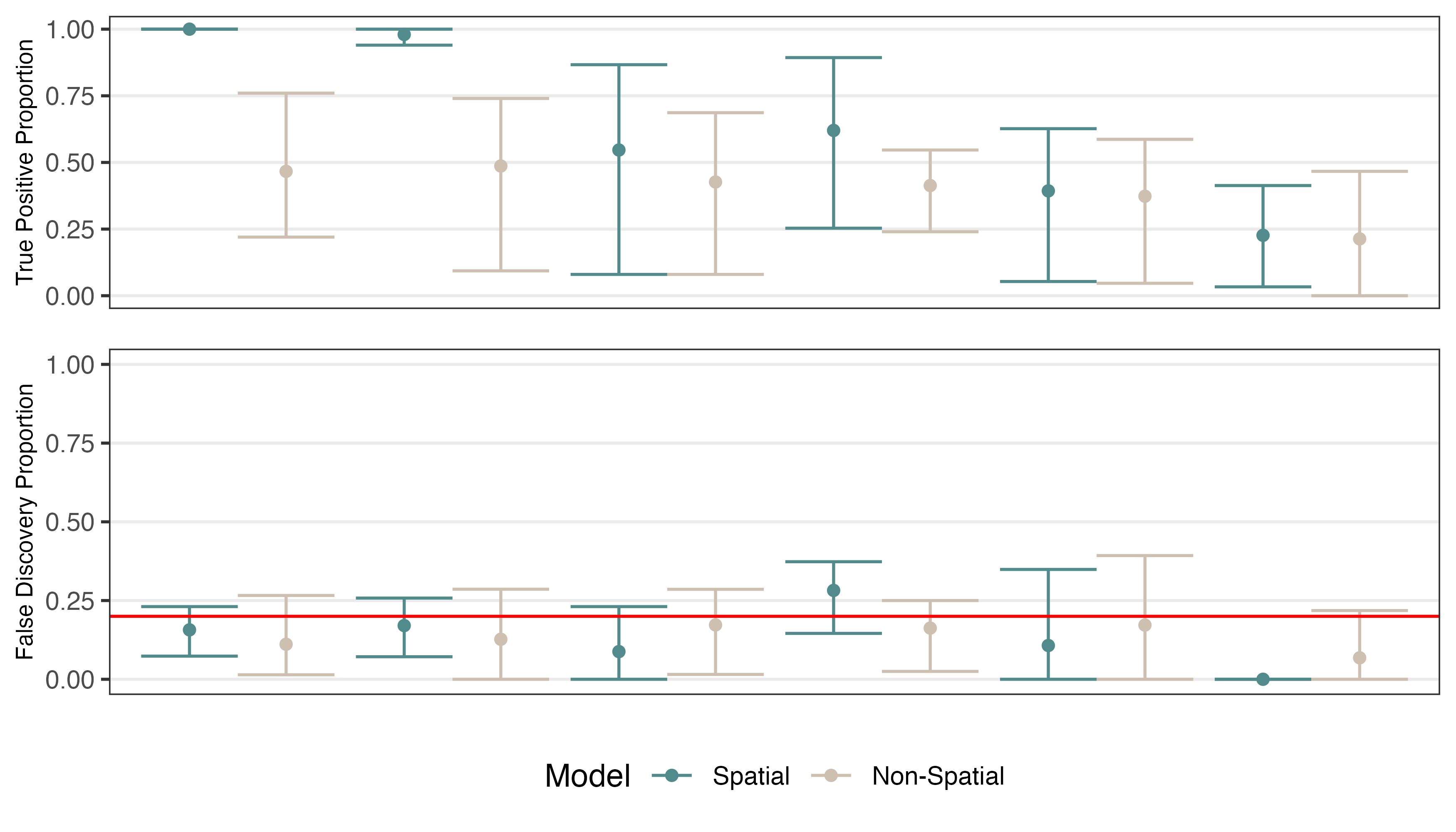} 
\caption{Means, $10$th, and $90$th quantiles of false discovery and true positive proportions. The true positive proportion is the number of causal SNPs correctly identified divided by $30$, the true number of relevant SNPs. The horizontal axis depicts the number of clusters of relevant SNPs and ranges from extreme (all relevant SNPs in one cluster) to no spatial structure (random position for all relevant SNPs). Results shown are for a noise to signal ratio of 5 (i.e., $\sigma^2_e=1$ in equation (\ref{BVSR_lin_mod}) and $u_{j}=0.2$ in equation (\ref{spike-and-slab prior})). Results are summarized across $10$ datasets for each clustering regime. All simulated datasets are for $n_u=20,000$ uncorrelated SNPs (no linkage disequilibrium) observed across $n=1,000$ unrelated (no population structure) individuals. The red horizontal line depicts the targeted false discovery proportion of $q=0.20$.} 
\label{errorbar_cluster}
\end{figure}

Power decreased in both models as the noise to signal ratio (NSR) increased (Figure \ref{errorbar_SNP}). Power was similar across models for datasets simulated without spatial structure, but the spatial model had greater power when the relevant SNPs were clustered (Figure \ref{errorbar_cluster}). As a result, mean power was only slightly lower for the non-spatial model for most NSRs, but the spatial model had a much higher upper bound for the two highest NSRs. Both models controlled FDR at the specified level for all NSRs and spatial clustering regimes.  

\begin{figure}[ht]
\centering
\includegraphics[scale=0.8]{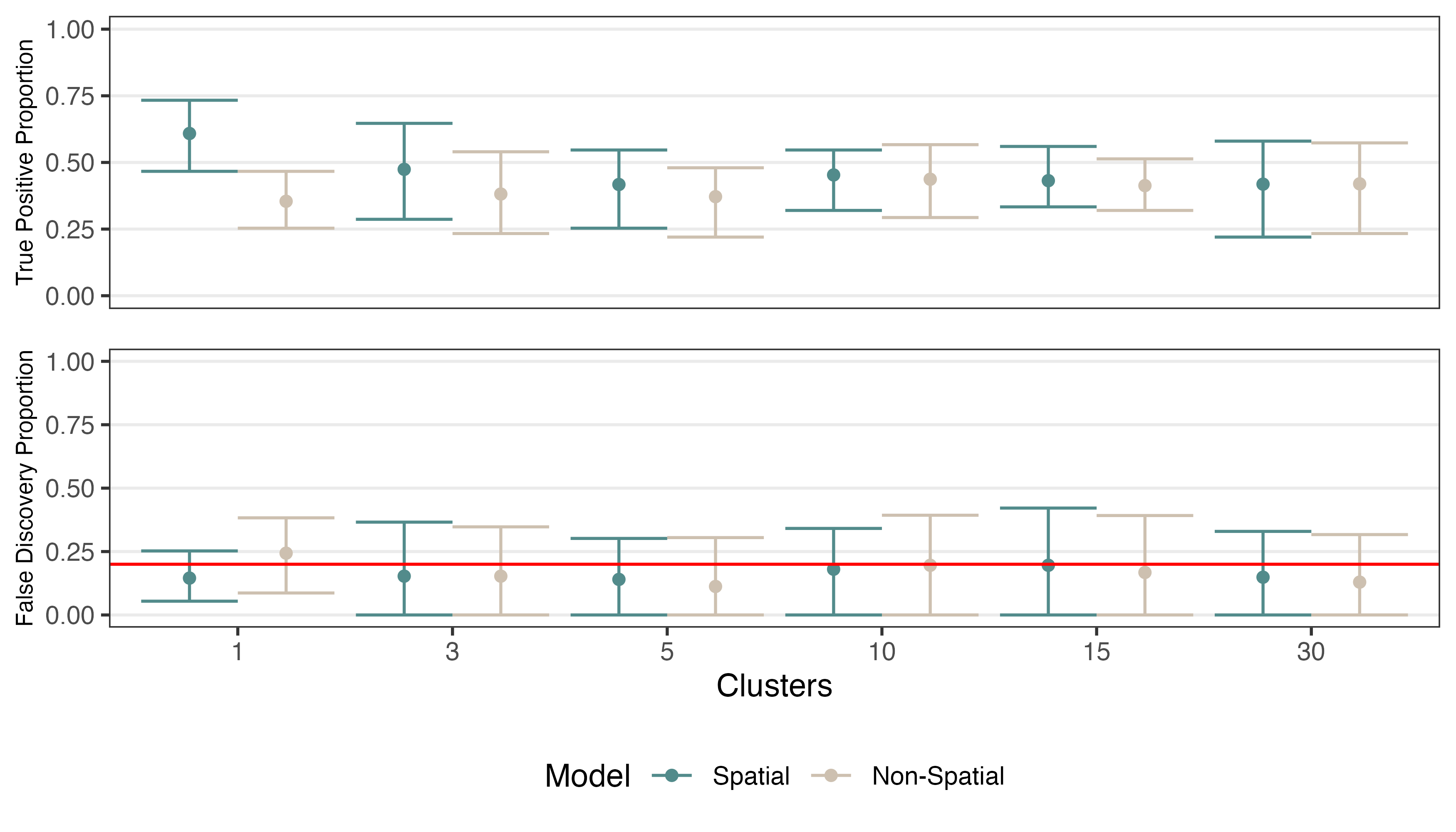} 
\caption{Means, $10$th, and $90$th quantiles of false discovery and true positive proportions. The true positive proportion is the number of causal SNPs correctly identified divided by $30$, the true number of relevant SNPs. The horizontal axis depicts the number of clusters of relevant SNPs and ranges from extreme (all relevant SNPs in one cluster) to no spatial structure (random position for all relevant SNPs). Results shown are for a noise to signal ratio of 1 (i.e., $\sigma^2_e=1$ in equation (\ref{BVSR_lin_mod}) and $u_{j}=1$ in equation (\ref{spike-and-slab prior})). Results are summarized across $27$ datasets for each clustering regime. All simulated datasets are for $n_u=20,000$ SNPs extracted from chromosomes $1$-$5$ of \textit{Arabidopsis thaliana} and $n=n_g=1,058$ related individuals with population structure mimicking \textit{Arabidopsis thaliana}. Linkage disequilibrium among SNPs was partially attenuated using the variable prunning procedure described in Web Supplement C. The red horizontal line depicts the targeted false discovery proportion of $q=0.20$.} 
\label{errorbar_cluster_Arabidopsis}
\end{figure}

Power decreased for both methods in the presence of linkage disequilibrium and population structure (Figure \ref{errorbar_cluster_Arabidopsis}). When SNPs are correlated, associations can become masked lowering power to detect relevant SNPs. For the datasets simulated without population structure and linkage disequilibrium, the relevant SNPs tended to be among the top $100$ SNPs with highest marginal association, but when we introduced  population structure and linkage disequilibrium, it was not uncommon to observe relevant SNPs with marginal associations that ranked higher than $10,000$. This trend is exacerbated with increased clustering of the relevant SNPs because linkage disequilibrium decays with genetic distance, and we observed that power decreased for the non-spatial model as we increased spatial dependence. The spatial model, on the other hand, increased in power as the relevant SNPs became more clustered. 

\begin{figure}[ht]
\centering
\includegraphics[scale=0.65]{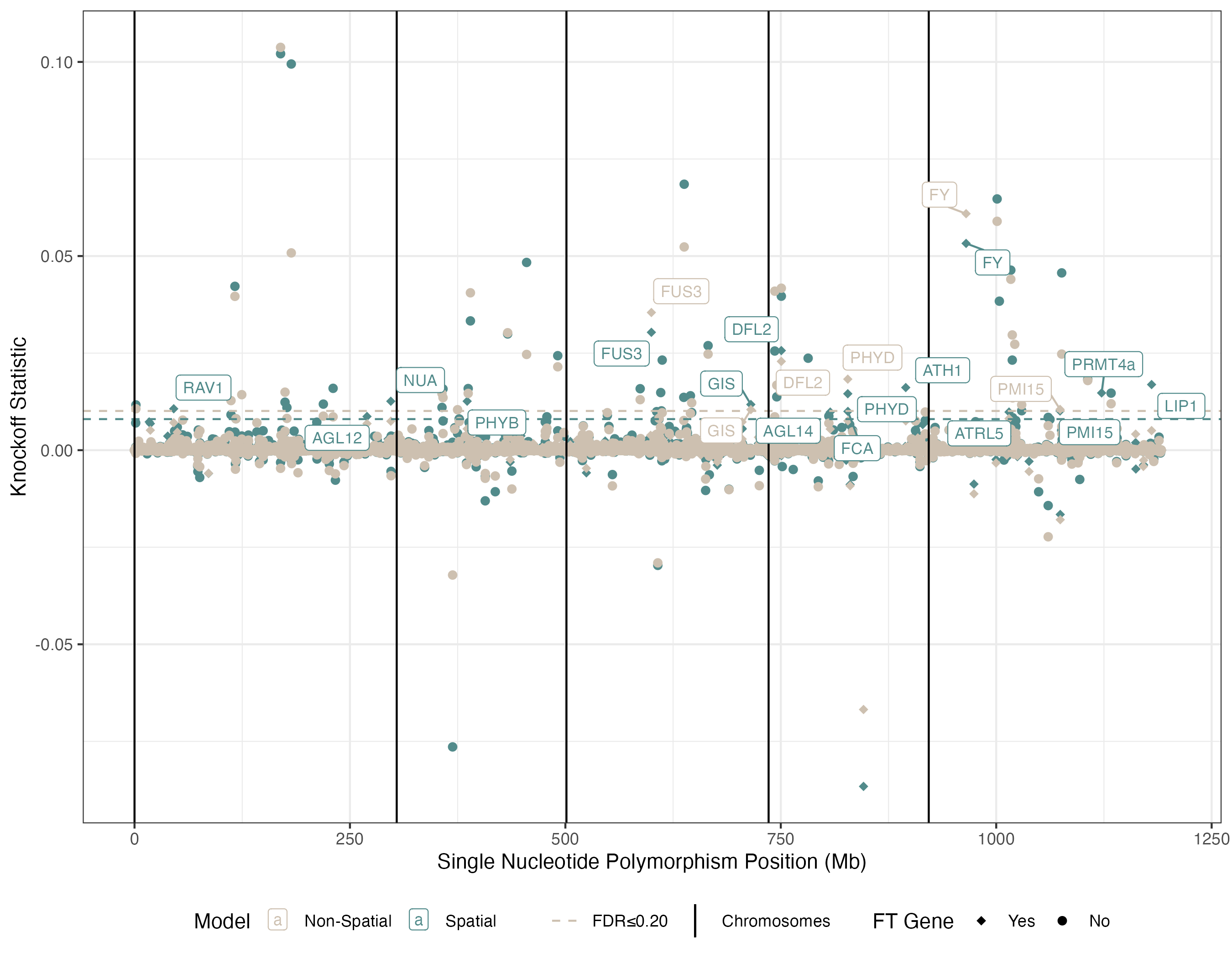} 
\label{knockoff_statistics_arabidopsis}
\caption{\textit{Arabidopsis thaliana} flowering time knockoff statistics for $n_u=558\text{,}321$ single nucleotide polymorphism (SNP) cluster representatives considered. Labels denote the $6$ and $16$ knockoff statistics for SNP clusters falling in buffers of flowering time genes selected in the non-spatial and spatial models, respectively. The five vertical lines depict the starting points of each chromosome. The dashed horizontal lines depict the threshold needed to obtain the targeted false discovery proportion of $q=0.20$. Shape indicates whether a SNP cluster fell within a $10$ kilobase buffer of one of the $282$ described flowering time genes in Brachi et al. (2010).} 
\end{figure}

Figure \ref{knockoff_statistics_arabidopsis} presents the knockoff statistics for $n_u=558\text{,}321$ SNP cluster representatives in \textit{Arabidopsis thaliana}. At a false discovery proportion threshold of $0.20$, we made $40$ and $66$ discoveries with the non-spatial and spatial models, respectively.  We identified SNPs likely tagging flowering time genes by noting whether any of the SNPs in their cluster fell within 10 kb, the estimated linkage disequilibrium rate of \textit{Arabidopsis thaliana} \citep{Kim2007}, of a described gene. The number of cluster representatives in flowering time gene buffers was $6$ and $16$ in the non-spatial and spatial models, respectively. Approximately $15.8\%$ of the $558\text{,}321$ SNP clusters in our subset had a cluster member that fell within the 10 kb buffer of one of the $282$ flowering time genes. Thus, in the non-spatial model, we selected roughly the proportion of SNP clusters tagging flowering time genes than we would have expected by chance ($15\%$), whereas in the spatial model, we selected a higher proportion of SNP representatives tagging flowering time genes ($24\%$). 

Both models selected SNPs within 10 kb buffers of the flowering time genes AT3G26790 (\textit{FUS3}), AT4G03400	(\textit{DFL2}), AT4G16250 (\textit{PHYD}) and AT5G13480 (\textit{FY}). One of the strongest associations estimated was for AT5G13480, which encodes a protein with similarity to yeast Pfs2p, an mRNA processing factor that influences flowering time \citep{Schmid2005}. Associations were also strong for AT3G26790, which regulates the hormones gibberellin and abscisic acid that can influence flowering time via their interactions with the transcription of the gene AT5G61850 (\textit{LEAFY}) \citep{Gazzarrini2004}. Many of the selected SNP clusters, while not directly linked to genes in the list described by \cite{Brachi2010}, are likely involved in the biochemical pathways regulating flowering time. For example, two additional discoveries made by the spatial model in chromosome $4$ had cluster members in the genes AT4G01650 and AT4G02120 (\textit{CTPS3}), which are expressed during many development stages including flowering \citep{Schmid2005}. 

\begin{figure}[ht]
\centering
\includegraphics[scale=0.65]{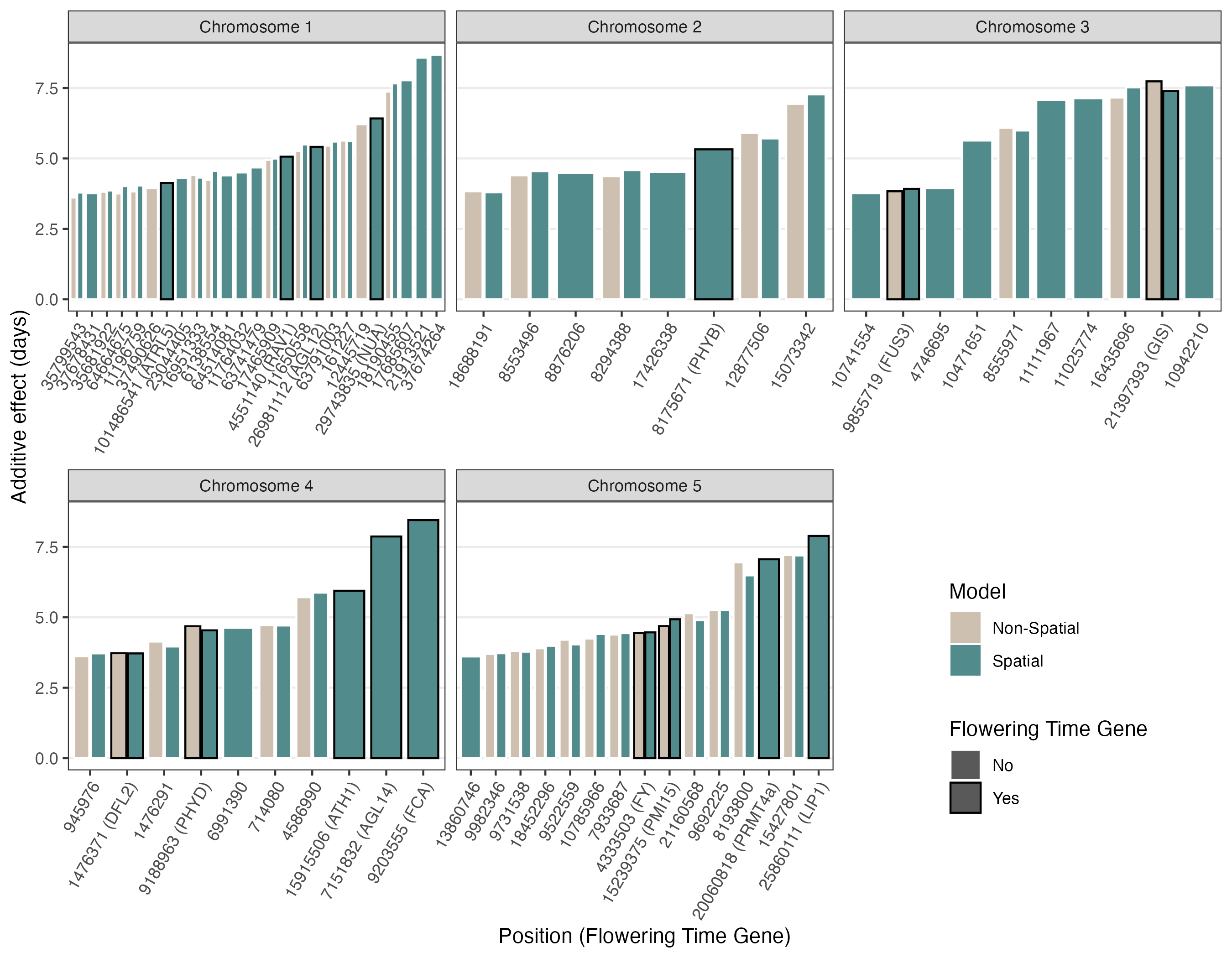} 
\caption{\textit{Arabidopsis thaliana} change in flowering time effects for $40$ and $66$ single nucleotide polymorphism (SNP) cluster representatives selected in non-spatial and spatial models, respectively. Effects represent the change in days to flowering induced by swapping two copies of the early flowering allele for two copies of the late flowering allele at a particular locus. Selected SNP clusters falling within a $10$ kilobase buffer one of the $282$ described flowering time genes in Brachi et al. (2010) are depicted by columns with a black outline and gene names are given in parenthesis.} 
\label{SNP_effects_arabidopsis}
\end{figure}

Figure \ref{SNP_effects_arabidopsis} gives the absolute value of the posterior means of $u_j$ for the subset of SNPs selected in each model. Because $\bm{G}$ is a variant matrix (i.e. binary), the absolute value of the posterior means of $u_j$ represent the expected change in days to flowering associated with replacing both copies of the early flowering allele with the late flowering allele at locus $j$. Effects varied from $2$-$9$ days and were similar for cluster representatives selected in both models. 

\section{Discussion}\label{Discussion_Section}

We developed a spatial BVSR model for improving power and controlling false discoveries. While we assessed the performance of the spatial BVSR model for GWAS, our method is more broadly applicable for high-dimensional variable selection and is catered for contexts in which the relevant variables have latent spatial structure (e.g., gene expression \citep{Stingo2011} or functional MRI studies \citep{Li2010}). Variable selection in GWAS is particularly challenging because SNP effects are often confounded with population structure. We demonstrated how to alleviate the confounding between SNP and population structure effects with restricted regression. Our restricted regression is appealing in GWAS because the SNP effects are estimated from the residuals of $\bm{y}$, or ``corrected'' trait, as proposed in previous methods \citep{Guan2011, Li2011, Lu2015, Sesia2021}. However, rather than correcting for population structure and then estimating SNP effects sequentially, we estimate all quantities simultaneously in a fully Bayesian model that accounts for all sources of uncertainty. 

In both the simulation study and analysis of \textit{Arabidopsis thaliana} flowering time, we showed that jointly modeling SNP inclusion probabilities can improve power when the relevant variants have latent spatial structure. The case study of \textit{Arabidopsis thaliana} flowering time also highlighted the advantages of the BVSR approach for identifying relevant variants as compared to standard single SNP approaches. In addition to being more powerful than the traditional single SNP approaches, the BVSR model also selected variables that varied considerably in their marginal associations. For example, among the $66$ SNPs selected in the spatial BVSR model, rankings based on marginal associations varied from $2$ to $2654$, with only $9$ SNP clusters selected from the top $100$ ranks. Traditional GWAS approaches tend to select only the lowest ranked SNPs. In the BVSR model, many of the low ranked SNPs are uncorrelated with the trait conditional on the hundreds of other SNPs included in the model, and hence go unselected. Because many of the low ranked SNPs are also likely relevant variants, we do not necessarily view this characteristic of BVSR as an advantage over traditional GWAS approaches, but rather a qualitative difference that supports using both methods together for describing complex genetic architectures. 

\cite{Zan2018} provided a comprehensive description of the genetic architecture underpinning \textit{Arabidopsis thaliana} flowering time and identified $33$ SNPs that collectively described $55.1\%$ of the total phenotypic variance in flowering time. \cite{Zan2018} first reduced the number of SNPs considered by screening for SNPs associated with the $282$ flowering time genes described in \cite{Brachi2010}. The joint effect and associations of the remaining SNPs was then estimated with a backward elimination association analysis with an adaptive FDR based threshold of $0.15$. The selected SNPs were congruent with flowering time genes with $11$ SNPs located within a flowering time gene. While the adherence of the SNPs selected with our spatial BVSR model to the flowering time genes was less strong, the selected SNPs still tended to be closer to flowering time genes than we would expect by chance. Furthermore, in cases where prior information is available regarding the location of genes, \cite{Fridley2009} noted performance could be improved by inflating the prior inclusion probabilities of these exonic SNPs by specifying a non-zero mean for $\bm{\alpha}$ in equation (\ref{alphaequation}). For the sake of comparison with the non-spatial BVSR model, we specified homogeneous prior inclusion probabilities for all SNPs in the \textit{Arabidopsis thaliana} analysis. Finally, not all relevant variants will be directly tied to described genes or exonic DNA \citep{Niu2019}, and the BVSR model provided a unique set of SNPs for further genomic investigation. 

Because Model-X knockoffs are stochastic, different runs of an algorithm can produce discrepancies in the selected variables based on the generated knockoff variables \citep{Ren2023}. In our analysis of \textit{Arabidopsis thaliana}, the lowest knockoff statistic for both models occurred in chromosome $4$ and is linked to a SNP cluster with representative in the flowering time gene AT4G20370 (\textit{TSF}). AT4G20370 is the twin sister of AT1G65480 (\textit{FT}), a previously identified flowering time gene for traditional GWAS approaches \citep{Alonso2016}. Just by chance, the knockoff copy of this SNP was more associated with flowering time than the original copy. \cite{Gu2021} suggested treating the knockoffs as random variables and sampling them directly in the MCMC algorithm. An advantage of this approach is that it can stabilize the knockoff filter by attenuating issues related to only using one realization of the knockoff variables. In principle, we could embed a Bayesian implementation of the HMM proposed by \cite{Sesia2019} within our BVSR and sample the knockoff SNPs each MCMC iteration, but this approach would be computationally infeasible for the large number of SNPs considered in most GWAS settings. We could take a derandomized knockoff approach and summarize the selected variables across multiple algorithm runs fit in parallel \citep{Ren2023}, but this would also be computationally infeasible for GWAS. 

The consequences of clustering for variable selection and multiple testing have been discussed at length \citep{Benjamini2007, Dai2016, Brzyski2017}. In GWAS, treating SNPs as independent units may not be appropriate if the goal is to discover genomic regions associated with a phenotype \citep{Wang2010, Lu2015, Brzyski2017, Candes2018, Sesia2019}. Variable selection within quantitative trait loci may not always be feasible because of high multicollinearity among clustered SNPs. We followed the method proposed by \cite{Candes2018} and \cite{Sesia2019} who suggested preprocessing the genotype matrix by first clustering SNPs with high mutual correlations and then choosing cluster representatives. Another option would be to include all SNPs but assign clusters to joint inclusion indicators as proposed by \cite{Lu2015} using group knockoffs \citep{Dai2016}. Our spatial BVSR model can be viewed as a more flexible version of this model that encourages nearby SNPs to have the same inclusion indicator as a result of the spatially structured inclusion probabilities. Modeling either the inclusion indicators or probabilities jointly has the potential to both improve power and reduce false discoveries because they can magnify true but negligible individual effects as well as dilute one-off spurious associations \citep{Benjamini2007, Lu2015, Brzyski2017}. 

Increased genomic and phenotypic data collection has highlighted the importance of methods for understanding the association between a response and an increasingly large number of predictors. Our proposed BVSR approach incorporates several recent advances in variable selection in one cohesive framework. Using restricted regression, we stabilized posterior computation for confounded factors that are generally estimated in a step-wise procedure. We highlighted FDR control with Model-X knockoffs in a setting of realistic complexity. We corroborated previous research that showed modeling the selection of variables jointly can improve performance \citep{Li2010}. Lastly, by combining reduced rank approximations and sampling strategies, we demonstrated the computational feasibility of Bayesian methods for high-dimensional variable selection. 

\textbf{Data Availability Statement}: The data that support the findings in this paper are available online from https://arapheno.1001genomes.org and https://aragwas.1001genomes.org. 

\textbf{Acknowledgements}: We thank several members from the BromeCast research network for their contributions to this manuscript. This research was funded by the National Science Foundation: NSF 222252, 1927177, 1927009, and 1927282.  


\bibliographystyle{biom}
\bibliography{bibliography.bib}

\end{document}


\title{\textbf{Web Supplement: Spatial Knockoff Bayesian Variable Section in Genome-Wide Association Studies}}

\maketitle

\vskip 2mm
\begin{center}
{Justin J. \textsc{Van Ee}$^{1}$} 
\let\thefootnote\relax\footnote{\baselineskip=7pt 1, 
Department of Statistics, Colorado State University, Fort Collins, CO}\textrm{\Large{, }}
{Diana \textsc{Gamba}$^{2}$}
\footnote{\baselineskip=7pt 2, Department of Biology, Pennsylvania State University, University Park, PA}\textrm{{, }}
{Jesse  R. \textsc{Lasky}$^{2}$}
\textrm{{, }}
\footnote{\baselineskip=7pt 3, Department of Wildland Resources and the Ecology Center, Utah State University, Logan, UT}
{Megan  L. \textsc{Vahsen}$^{3}$}
\footnote{\baselineskip=7pt 4,  Department of Statistics and Data Sciences, The University of Texas at Austin, Austin, TX} 
\textrm{{, }}
{Mevin  B. \textsc{Hooten}$^{4}$}
\end{center}

\appendix
\singlespacing

\section{Markov chain Monte Carlo Sampling Strategies}\label{MCMC_strategies}

We obtained a posterior sample for all quantities using Markov chain Monte Carlo (MCMC). The crux of making MCMC computationally feasible in this setting was to avoid sampling the parameters with dimension $n_u$ each MCMC iteration. These parameters include $\bm{\nu}$, $\bm{\delta}$, $\tilde{\bm{\delta}}$, $\bm{u}$, and $\tilde{\bm{u}}$. The BVSR approaches obviated repeated sampling of all elements in $\bm{u}$ and $\tilde{\bm{u}}$ because, we have $u_j=\tilde{u}_j=0$ whenever $\nu_j=0$. 

Updating the individual elements of the SNP inclusion indicators, $\bm{\nu}$, $\bm{\delta}$, and $\tilde{\bm{\delta}}$, is less straightforward, and we apply several strategies to improve mixing and reduce computational time. We updated the SNP inclusion indicators using add, delete, and swap steps randomly selected with probabilities $(0.45, 0.45, 0.1)$, respectively \citep{Guan2011}. With probability $p_{\text{SW}}$, we compounded between $2$-$n_{\text{SW}}$ add and delete steps to form a long-range proposal \citep{Guan2011}, where $p_{\text{SW}}$ and $n_{\text{SW}}$ are tuning parameters and their optimal values depend on the setting and data. This technique, referred to as ``small-world proposal,'' improves the theoretical convergence rate of the MCMC scheme \citep{Guan2007}.  We also facilitated mixing for the knockoff indicators by uniformly selecting an included SNP $j$ and proposing $(\delta_j,\tilde{\delta}_j)\rightarrow(\tilde{\delta}_j, \delta_j)$ each MCMC iteration.

In an add step, we assigned priority to the SNPs with the greatest marginal associations by proposing to include them at a higher frequency. Specifically, we ranked both the original and knockoff copies of each SNP from $1$ to $n_u$ based on their individual $p$-values from the linear regression models,
\begin{align}
    \bm{y}&\sim\mathcal{N}(\bm{R}\bm{\theta}+\bm{X}\bm{\beta}+\bm{Z}\bm{K}_ju_j, \sigma^2_{e}\bm{I}), \label{original_ranks}\\
    \bm{y}&\sim\mathcal{N}(\bm{R}\bm{\theta}+\bm{X}\bm{\beta}+\bm{Z}\tilde{\bm{K}}_j\tilde{u}_j, \sigma^2_{e}\bm{I}),
\end{align}
respectively. When adding a SNP to the model, we first drew a $\text{Bernoulli}(0.5)$ random variable to decide between adding the original or knockoff copy of the SNP. We then drew a random rank, $r$, from a mixture of geometric and discrete uniform distributions truncated to $n_u$ and selected the corresponding copy of the SNP with rank $r$ \citep{Guan2011}. For the delete steps, SNPs are selected uniformly. Lastly, in a swap, we simultaneously proposed to add a SNP not currently included in the model using the ranked proposal described above while uniformly selecting one of the included SNPs to delete. 

The knockoff statistics, $w_j=\delta_j-\tilde{\delta}_j$, are prone to high sampling variance because only a tiny fraction of the $n_u$ SNPs are included in the model (i.e., $\nu_j=1$) for a given MCMC iteration. Hence, the vast majority of samples in a Markov chain for $w_j$ are zero. The posterior mean of the $j$th knockoff statistics is given by $\mathbb{E}(w_j|\bm{y})=\mathcal{P}(\delta_j=1|\bm{y})-\mathcal{P}(\tilde{\delta}_j=1|\bm{y})$, which is the difference of the posterior inclusion probabilities for the original and knockoff variables. We estimated both posterior inclusion probabilities using their Rao–Blackwellized estimate \citep{Casella1996} to increase sampling efficiency \citep{Guan2011, Barker2013}. We provide a derivation of the Rao–Blackwellized estimates of $\mathcal{P}(\delta_j=1|\bm{y})$ and $\mathcal{P}(\tilde{\delta}_j=1|\bm{y})$ in Web Supplement D. We sampled $\bm{\beta}$, $\bm{\theta}$, $\bm{u}$, $\tilde{\bm{u}}$, $\sigma^2_e$, $\sigma^2_{\theta}$,  and $\tau$ from their full-conditional distributions, and obtained samples of $\sigma^2_a$, $\pi$, $\bm{\alpha}$, and $\mu_{\pi}$ using a Gaussian random walk Metropolis-Hastings algorithm with automatic tuning \citep{Shaby2010}. 

\section{Simulation Study}

\subsection{Non-Spatial Model}\label{SS_simulationstudy}

\begin{align}
    \bm{y} &\sim\mathcal{N}(\bm{X}\bm{\beta}+\bm{G}\bm{u}+\tilde{\bm{G}}\tilde{\bm{u}}, \sigma^2_{e}\bm{I}), \\
    (u_j, \tilde{u}_j) & \sim (\delta_j, \tilde{\delta}_j)\mathcal{N}(0,\sigma^2_{a}), \\
    (\delta_j, \tilde{\delta}_j) & \sim \begin{cases}
    \text{Categorical}(0.5,0.5), & \text{for $\nu_j=1$} \\
    (0,0), & \text{for $\nu_j=0$} 
    \end{cases}, \\
    \nu_j & \sim\text{Bernoulli}(\pi), \\
    \sigma^2_a(h, \bm{\delta}, \tilde{\bm{\delta}}, \bm{s}^2, \tilde{\bm{s}}^{2})&=\frac{h}{1-h}\frac{1}{\sum_{j:\delta_j=1}s_j^2+\sum_{j:\tilde{\delta}_j=1}\tilde{s}_j^2}, \\
    h&\sim\mathcal{U}(0,1), \\
    \log(\pi)&\sim\mathcal{U}(\log(1/n_u),\log(100/n_u)), \\
    \bm{\beta}&\sim\mathcal{N}(\bm{0},100^2\bm{I}), \\
    \sigma_e^2&\sim\mathcal{IG}(0.001, 1000).
\end{align}

\subsection{Spatial Model}

\begin{align}
    \bm{y} &\sim\mathcal{N}(\bm{X}\bm{\beta}+\bm{G}\bm{u}+\tilde{\bm{G}}\tilde{\bm{u}}, \sigma^2_{e}\bm{I}), \\
    (u_j, \tilde{u}_j) & \sim (\delta_j, \tilde{\delta}_j)\mathcal{N}(0,\sigma^2_{a}), \\
    (\delta_j, \tilde{\delta}_j) & \sim \begin{cases}
    \text{Categorical}(0.5,0.5), & \text{for $\nu_j=1$} \\
    (0,0), & \text{for $\nu_j=0$} 
    \end{cases}, \\
    \nu_j & \sim\text{Bernoulli}(\pi_j), \\
    \sigma^2_a(h, \bm{\delta}, \tilde{\bm{\delta}}, \bm{s}^2, \tilde{\bm{s}}^{2})&=\frac{h}{1-h}\frac{1}{\sum_{j:\delta_j=1}s_j^2+\sum_{j:\tilde{\delta}_j=1}\tilde{s}_j^2}, \\
    h&\sim\mathcal{U}(0,1), \\
    \text{logit}(\bm{\pi}) & =\mu_{\pi}\bm{1}+\bm{B}\bm{\alpha}, \label{B_sim} \\
    \bm{\alpha} & \sim\mathcal{N}(\bm{0}, \tau\bm{B}'\bm{L}(\rho)\bm{B}), \\   \bm{\beta}&\sim\mathcal{N}(\bm{0},100^2\bm{I}), \\
    \mu_{\pi} & \sim\mathcal{N}(-7, 2.25^2), \\ 
    \tau&\sim\mathcal{IG}(0.001, 1000), \\
    \sigma_e^2&\sim\mathcal{IG}(0.001, 1000).
\end{align}

The prior on $\mu_{\pi}$ implied an expectation of $173$ relevant SNPs with $95\%$ prior credible interval $(0, 1301)$ for the $n_u=20\text{,}000$ considered in the simulation study. 

\subsection{Data Simulation}

We performed a simulation study to measure the performance of our spatial spike-and-slab model for identifying causal SNPs. In each simulation, we simulated $n=1000$ phenotypes from $n_g=1000$ unrelated genotypes using the linear model $\bm{y}\sim\mathcal{N}(\bm{X}\bm{\beta}+\bm{Z}\bm{G}\bm{u}+\bm{Z}\tilde{\bm{G}}\tilde{\bm{u}}, \sigma^2_{e}\bm{I})$. The matrix $\bm{X}$ included an intercept, and two continuous covariates. We set $n_u=20,000$ and specified $30$ relevant (i.e., $u_j\ne 0$) SNPs. Across simulations, we varied the degree of spatial structure in the SNP inclusion indicators by manipulating the number of SNP clusters. The spatial structure varied from extreme (all $30$ relevant SNPs in one randomly located cluster) to none (random location for each relevant SNP). We set $\sigma^2_e=1$ but varied the effect size of the causal SNPs so that the noise to signal ratio varied between $1$-$6$ across simulations. For each signal-spatial structure combination, we simulated $10$ datasets, for a total of $360$ simulated datasets.  For the spatial model, we assumed a simple neighborhood structure in which each SNP was only adjacent to the two SNPs preceding and following it.

To assess the performance of our model in the presence of population structure and linkage disequilibrium, we also simulated datasets using subsets of the \textit{Arabidopsis thaliana} genotype matrix. Specifically, we partitioned the pruned \textit{Arabidopsis thaliana} genotype matrix described in Web Supplement (\ref{data_arabidopsis}) into $27$ non-overlapping  genotype matrices, each with $n_u=20,000$ and $n=n_g=1058$. For every partition, we simulated one dataset for each of six different clustering regimes and specified $30$ relevant SNPs with a noise to signal ratio of $1$ for all simulated datasets. We considered several neighborhood matrices based on the spatial arrangement of SNPs in the extracted genotype matrices. Overall, we found that the neighborhood matrices derived from the distance matrix of SNPs resulted in worse performance than from matrices based on the simple neighborhood structure described above. All results displayed in the manuscript are for simple neighborhood matrices. 

\subsection{Markov Chain Monte Carlo Details}

For the add steps in the algorithm used to update $\nu_j$, $\delta_j$, and $\tilde{\delta}_j$, we drew the rank, $r$, of the proposed SNP from the mixture distribution $r\sim \big(0.7\times\text{Geometric}(0.05, n_u)+0.3\times\text{Discrete Uniform}(0,n_u)\big)$, where $\text{Geometric}(0.05, n_u)$ denotes a truncated Geometric distribution with mean parameter $20$ and upper limit $n_u$. We set $n_{\text{SW}}=10$ and $p_{\text{SW}}=0.3$ such that for roughly $30\%$ of MCMC iterations we compounded between $2$-$10$ add and delete steps to form a small world proposal \citep{Guan2007}. This resulted in an acceptance rate of approximately $25\%$ for small world proposals. In the simulation study, we fit both the non-spatial and spatial models for $1$ million iterations, discarded the first $500$ thousand iterations as burn-in, and thinned the remaining sample to $200$ for calculating all posterior quantities of interest. Run times were $16$ and $7$ minutes for the non-spatial and spatial models, respectively, on a machine with a 3.5 GHz processor and 96 GB RAM. The matrix $\bm{B}$, equation (\ref{B_sim}), included the first $25$ singular vectors from the singular value decomposition (SVD) of the simulated neighborhood matrix for the purely synthetic datasets and the first $100$ singular vectors for datasets simulated using subsets of the \textit{Arabidopsis thaliana} genotype matrix. 

\section{Analysis of \textit{Arabidopsis thaliana} Flowering Time}

\subsection{Non-Spatial Model}

\begin{align}
    \bm{y} &\sim\mathcal{N}(\bm{R}\bm{\theta}+\bm{X}\bm{\beta}+\bm{K}\bm{u}+\tilde{\bm{K}}\tilde{\bm{u}}, \sigma^2_{e}\bm{I}), \\
    (u_j, \tilde{u}_j) & \sim (\delta_j, \tilde{\delta}_j)\mathcal{N}(0,\sigma^2_{a}), \\
    (\delta_j, \tilde{\delta}_j) & \sim \begin{cases}
    \text{Categorical}(0.5,0.5), & \text{for $\nu_j=1$} \\
    (0,0), & \text{for $\nu_j=0$} 
    \end{cases}, \\
    \nu_j & \sim\text{Bernoulli}(\pi), \\
    \sigma^2_a(h, \bm{\delta}, \tilde{\bm{\delta}}, \bm{s}^2, \tilde{\bm{s}}^{2})&=\frac{h}{1-h}\frac{1}{\sum_{j:\delta_j=1}s_j^2+\sum_{j:\tilde{\delta}_j=1}\tilde{s}_j^2}, \\
    h&\sim\mathcal{U}(0,1), \\
    \log(\pi)&\sim\mathcal{U}(\log(1/n_u),\log(10/n_u)), \label{FT_U}\\
    \bm{\beta}&\sim\mathcal{N}(\bm{0},100^2\bm{I}), \\
    \bm{\theta}&\sim\mathcal{N}(\bm{0}, \sigma^2_{\theta}\bm{I}), \\
    \sigma_{\theta}^2&\sim\mathcal{IG}(0.001, 1000), \\
    \sigma_e^2&\sim\mathcal{IG}(0.001, 1000),
\end{align}
where $\bm{K}\bm{u}+\tilde{\bm{K}}\tilde{\bm{u}}=(\bm{I}-\bm{R}(\bm{R}'\bm{R})^{-1}\bm{R}')\left(\bm{G}\bm{u}+\tilde{\bm{G}}\tilde{\bm{u}}\right)$.

\subsection{Spatial Model}\label{SP_FT}

\begin{align}
    \bm{y} &\sim\mathcal{N}(\bm{R}\bm{\theta}+\bm{X}\bm{\beta}+\bm{K}\bm{u}+\tilde{\bm{K}}\tilde{\bm{u}}, \sigma^2_{e}\bm{I}), \\
    (u_j, \tilde{u}_j) & \sim (\delta_j, \tilde{\delta}_j)\mathcal{N}(0,\sigma^2_{a}), \\
    (\delta_j, \tilde{\delta}_j) & \sim \begin{cases}
    \text{Categorical}(0.5,0.5), & \text{for $\nu_j=1$} \\
    (0,0), & \text{for $\nu_j=0$} 
    \end{cases}, \\
    \nu_j & \sim\text{Bernoulli}(\pi_j), \\
    \sigma^2_a(h, \bm{\delta}, \tilde{\bm{\delta}}, \bm{s}^2, \tilde{\bm{s}}^{2})&=\frac{h}{1-h}\frac{1}{\sum_{j:\delta_j=1}s_j^2+\sum_{j:\tilde{\delta}_j=1}\tilde{s}_j^2}, \\
    h&\sim\mathcal{U}(0,1), \\
    \text{logit}(\bm{\pi}_{k}) & ={\mu_{\pi}}_k\bm{1}+\bm{B}_k\bm{\alpha}_k,  && \text{ for $k=1,\dots,5$}\label{FT_B} \\
    \bm{\alpha}_k & \sim\mathcal{N}(\bm{0}, \tau\bm{B}_k'\bm{L}(\rho)\bm{B}_k), \\   \bm{\beta}&\sim\mathcal{N}(\bm{0},100^2\bm{I}), \\
    \mu_{\pi} & \sim\mathcal{N}(-12, 2.25^2), \\ 
    \tau&\sim\mathcal{IG}(0.001, 1000), \\
    \bm{\theta}&\sim\mathcal{N}(\bm{0}, \sigma^2_{\theta}\bm{I}), \\
    \sigma_{\theta}^2&\sim\mathcal{IG}(0.001, 1000), \\
    \sigma_e^2&\sim\mathcal{IG}(0.001, 1000).
\end{align}

The prior on $\mu_{\pi}$ implied an expectation of $43$ relevant SNPs with $95\%$ prior credible interval $(0, 282)$ for the $n_u=558\text{,}321$ considered in the \textit{Arabidopsis thaliana} flowering time analysis. 

\subsection{Data Acquisition and Preprocessing}\label{data_arabidopsis}

We downloaded mean flowering times, averaged across accessions, at $10^\circ$C for $1058$ wild collected lines of \textit{Arabidopsis thaliana} from AraPheno \citep{Seren2016}, and obtained the genotype matrix of $7,363,011$ SNPs for the $1058$ lines from the AraGWAS Catalog  \citep{Togninalli2018}. Most lines have nearly completely homozygous genomes as a result of natural inbreeding. The genotype matrix is coded as $0$ and $1$ to represent homozygous for the reference and alternate alleles, respectively \citep{Jaegle2023}. 

High correlations among neighboring SNPs because of linkage disequilibrium decreases power to detect relevant variants in GWAS \citep{Brzyski2017, Candes2018, Zhu2021}. Grouping SNPs into clusters and performing selection on the groups or cluster representatives has been shown to improve performance \citep{Wang2010, Fridley2011, Lu2015, Candes2018, Sesia2019}. We followed the variable pruning procedure advocated by \cite{Candes2018} and \cite{Sesia2019}. We hierarchically clustered the SNPs using their absolute correlation as a measure of similarity, and cut the dendrogram at the height such that collections of SNPs having mutual correlations of $0.5$ or more were classified into clusters. We then used $20\%$ of the observations of flowering time to calculate marginal t-tests using equation (\ref{original_ranks}) for each SNP within clusters and choose the SNP with the lowest $p$-value as the cluster representative. 

The variable pruning procedure reduced the total number of SNPs considered to $n_u=558\text{,}321$. We generated knockoffs for the pruned genotype using the algorithm of \cite{Sesia2019}. Because the genotype matrix is binary, we used the haplotype implementation of \texttt{FastPhase} and \texttt{SNPknock} even though \textit{Arabidopsis thaliana} is diploid. Following \cite{Candes2018}, we set the rows of the knockoff genotype matrix corresponding to the observations used for determining the cluster representatives to their original values to ensure they met the exchangeability lemma described in \cite{Barber2015}. 

We calculated the identity-by-state (IBS) kinship matrix \citep{Stevens2011} using the pruned genotype matrix with \texttt{R} package \texttt{statgenGWAS} and let $\bm{R}$ equal the top five singular vectors from the kinship matrix. Similar patterns were explained by the SVD from genotype matrix, but the proportion of variance explained by each individual singular vector was much smaller than the components obtained from the kinship matrix. The matrix $\bm{X}$ included a single column of ones. Based on the findings of our simulation study, we specified a simple neighborhood structure for the proximity matrix, $\bm{A}$, in all five chromosomes.  

\subsection{Markov Chain Monte Carlo Details}

We ran the MCMC algorithms for the spatial and non-spatial model for $3$ million iterations each. Run times were $2$ and $12$ hours, respectively. Of the $550$K SNPs considered, over $3\text{,}000$ fell within one of the $282$ flowering time genes described in \cite{Brachi2010}. Because the number of relevant SNPs is likely much larger than can be accurately estimated with $1058$ genotypes \citep{Brachi2010} and to reduce computational burden, we regularized the number of SNPs included in the model by truncating the allowable sparsity at an upper limit. Specifically, in the non-spatial model, we let $b=\log(10/n_u)$ in equation (\ref{FT_U}) and, for the spatial model, we imposed the joint linear restriction $\bar{\pi}\le\log(10/n_u)$ on the parameters $\mu_{\pi}$ and $\bm{\alpha}$ in equation (\ref{FT_B}), where $\bar{\pi}$ denotes the arithmetic mean of $\bm{\pi}$. These restrictions held the number of included SNPs, the sum of original and knockoff copies, between $50$-$200$. 

For the add steps in the algorithm used to update $\nu_j$, $\delta_j$, and $\tilde{\delta}_j$, we drew the rank, $r$, of the proposed SNP from the mixture distribution $r\sim \big(0.2\times\text{Geometric}(0.05, n_u)+0.8\times\text{Discrete Uniform}(0,n_u)\big)$, where $\text{Geometric}(0.05, n_u)$ denotes a truncated Geometric distribution with mean parameter $20$ and upper limit $n_u$. We set $n_{\text{SW}}=15$ and $p_{\text{SW}}=0.3$ such that for roughly $30\%$ of MCMC iterations we compounded between $2$-$15$ add and delete steps to form a small world proposal \citep{Guan2007}. This resulted in an acceptance rate of approximately $10\%$ for small world proposals. The matrix $\bm{B}_k$, equation (\ref{FT_B}), included the first $100$ singular vectors from the SVD of the neighborhood matrix of pruned SNPs for chromosome $k$ of \textit{Arabidopsis Thaliana}. 

\section{Rao–Blackwellized Estimates of Knockoff Statistics}

Implementing the described BVSR models with MCMC results in high sampling variance for the knockoff statistics, $w_j=\delta_j-\tilde{\delta}_j$, because only a tiny proportion ($\approx 0.1\%$ for the flowering time analysis) of the $n_u$ SNPs are included in the model (i.e., $\nu_j=1$) for a given MCMC iteration. We derived the Rao–Blackwellized estimates \citep{Casella1996} of the posterior mean of the knockoff statistics $\mathbb{E}(w_j|\bm{y})=\mathcal{P}(\delta_j=1|\bm{y})-\mathcal{P}(\tilde{\delta}_j=1|\bm{y})$. Without loss of generality, we show derivations for the estimates of the spatial model in the flowering time (\ref{SP_FT}), noting the derivations are simpler for the other three. 

Observe 
\begin{align}
    \mathbb{E}(w_j|\bm{y}) & =\mathcal{P}(\delta_j=1|\bm{y})-\mathcal{P}(\tilde{\delta}_j=1|\bm{y}), \\
    & = \mathcal{P}(\delta_j=1, \nu_j=1|\bm{y})-\mathcal{P}(\tilde{\delta}_j=1, \nu_j=1|\bm{y}), \\
    & = \left(\mathcal{P}(\delta_j=1|\nu_j=1,\bm{y})-\mathcal{P}(\tilde{\delta}_j=1| \nu_j=1,\bm{y})\right)\mathcal{P}(\nu_j=1|\bm{y}), \\
    & = \left(\mathcal{P}(\delta_j=1|\nu_j=1,\bm{y})-\mathcal{P}(\delta_j=0| \nu_j=1,\bm{y})\right)\mathcal{P}(\nu_j=1|\bm{y}), \\
    & = \left(2\mathcal{P}(\delta_j=1|\nu_j=1,\bm{y})-1\right)\mathcal{P}(\nu_j=1|\bm{y}). 
\end{align}
The Rao-Blackwellized estimate of $\mathbb{E}(w_j|\bm{y})$ is 
\begin{align}\label{Rao_Blackwell_Estimate}
    \frac{1}{n_{\text{MCMC}}}\sum_{i=1}^{n_{\text{MCMC}}} \left(2\mathcal{P}(\delta_j=1|\nu_j=1,\bm{y},\bm{\gamma}_{-j}^{(i)})-1\right)\mathcal{P}(\nu_j=1|\bm{y},\bm{\gamma}_{-j}^{(i)})
\end{align}
where $\bm{\gamma}_{-j}^{(i)}=(\bm{\nu}_{-j}', \bm{u}_{-j}', \tilde{\bm{u}}_{-j}', \bm{\delta}_{-j}', \tilde{\bm{\delta}}_{-j}', \bm{\beta}', \bm{\theta}', \bm{\alpha}', \mu_{\pi}, \tau, \sigma^2_e,\sigma^2_{\theta})$ denotes the value of the vector of parameters at MCMC iteration $i$, and we use the subscript ``${-j}$'' to denote a vector excluding its $j$th element.  
Let 
\begin{align*}
    \eta_j&=[\bm{y}|\delta_j=1,\bm{\gamma}_{-j}^{(i)}][\bm{u}_{-j}|\delta_j=1,\bm{\gamma}_{-j}^{(i)}][\tilde{\bm{u}}_{-j}|\delta_j=1,\bm{\gamma}_{-j}^{(i)}], \\
    \tilde{\eta}_j&=[\bm{y}|\tilde{\delta}_j=1,\bm{\gamma}_{-j}^{(i)}][\bm{u}_{-j}|\tilde{\delta}_j=1,\bm{\gamma}_{-j}^{(i)}][\tilde{\bm{u}}_{-j}|\tilde{\delta}_j=1,\bm{\gamma}_{-j}^{(i)}],
\end{align*}
where we have conditioned each probability distribution on the entire vector of parameters $\bm{\gamma}_{-j}^{(i)}$ for brevity, but note that many of the parameters are independent conditional on the others. It follows that 
\begin{align}
    \mathcal{P}(\delta_j=1|\nu_j=1,\bm{y},\bm{\gamma}_{-j}^{(i)})=\frac{\eta_j}{\eta_j
    +\tilde{\eta}_j}.
\end{align}
By similar manipulations, we have 
\begin{align}
    \mathcal{P}(\nu_j=1|\bm{y},\bm{\gamma}_{-j}^{(i)})=\frac{\lambda_j}{\lambda_j
    +1},
\end{align}
where 
\begin{align}\label{lambdaj}
    \lambda_j=\frac{\frac{1}{2}\left(\eta_j+\tilde{\eta}_j\right)[\nu_j=1|\bm{\gamma}_{-j}^{(i)}]}{[\bm{y}|\nu_j=0,\bm{\gamma}_{-j}^{(i)}][\bm{u}_{-j}|\nu_j=0,\bm{\gamma}_{-j}^{(i)}][\tilde{\bm{u}}_{-j}|\nu_j=0,\bm{\gamma}_{-j}^{(i)}][\nu_j=0|\bm{\gamma}_{-j}^{(i)}]}.
\end{align}
We derive analytical expressions for all quantities in equation (\ref{lambdaj}) below. 

First, observe that 
\begin{align}
    \frac{[\nu_j=1|\bm{\gamma}_{-j}^{(i)}]}{[\nu_j=0|\bm{\gamma}_{-j}^{(i)}]}=\frac{\pi_j}{1-\pi_j}.
\end{align}
The terms $[\bm{u}_{-j}|\nu_j=0,\bm{\gamma}_{-j}^{(i)}]$, $[\tilde{\bm{u}}_{-j}|\nu_j=0,\bm{\gamma}_{-j}^{(i)}]$, $[\bm{u}_{-j}|\delta_j=1,\bm{\gamma}_{-j}^{(i)}]$, $[\tilde{\bm{u}}_{-j}|\delta_j=1,\bm{\gamma}_{-j}^{(i)}]$, $[\bm{u}_{-j}|\tilde{\delta}_j=1,\bm{\gamma}_{-j}^{(i)}]$, and $[\tilde{\bm{u}}_{-j}|\tilde{\delta}_j=1,\bm{\gamma}_{-j}^{(i)}]$ are all independent univariate normal densities of the form 
\begin{align}
    [\bm{u}_{-j}|\nu_j=0,\bm{\gamma}_{-j}^{(i)}]=\prod_{i\ne j}\mathcal{N}\left(u_{i}|0, \sigma^2_a(h, \bm{\delta}, \tilde{\bm{\delta}}, \bm{s}, \tilde{\bm{s}})\right), 
\end{align}
that arise because the value of the variance $\sigma^2_a$ depends on the indicators  $\bm{\delta}$ and $\tilde{\bm{\delta}}$. For $[\bm{y}|\delta_j=1,\bm{\gamma}_{-j}^{(i)}]$, we have 
\begin{align}
    & [\bm{y}|\delta_j=1,\bm{\gamma}_{-j}^{(i)}] \\
    &=\int[\bm{y},u_j|\delta_j=1,\bm{\gamma}_{-j}^{(i)}]du_j \\
    &=\int \left(2\uppi\sigma^2_e\right)^{-n/2}\left(2\uppi\sigma^2_a\right)^{-1/2}\exp{\left\{-\frac{1}{2\sigma^2_e}(\bm{R}-\bm{K}_j u_j)'(\bm{R}-\bm{K}_j u_j)-\frac{1}{2\sigma^2_a}u_j^2\right\}}du_j,
\end{align}
where $\bm{\epsilon}=\bm{y}-(\bm{X}\bm{\beta}+\bm{R}\bm{\theta}+\bm{K}_{\bm{\delta}-j}\bm{u}_{\bm{\delta}-j}+\tilde{\bm{K}}_{\tilde{\bm{\delta}}-j}\tilde{\bm{u}}_{\tilde{\bm{\delta}}-j})$, $\bm{\delta}-j$ denotes the vector $\bm{\delta}$ with $j$th element set to $0$. Integrating out $u_j$, we obtain,
\begin{align}\label{originallikelihood}
    [\bm{y}|\delta_j=1,\bm{\gamma}_{-j}^{(i)}]
    =\left(2\uppi\sigma^2_e\right)^{-n/2}\left(\sigma^2_a\right)^{-1/2}\left(a_{\star}\right)^{1/2}\exp{\left\{-\frac{1}{2\sigma^2_e}\bm{\epsilon}'\bm{\epsilon}+\frac{1}{2 a_{\star}}b_{\star}^2\right\}},
\end{align}
where 
\begin{align}
    a_{\star}&=\left(\frac{\bm{K}_j'\bm{K}_j}{\sigma^2_e}+\frac{1}{\sigma^2_a}\right)^{-1}, \\
    b_{\star}&=\frac{1}{\sigma^2_e}\bm{\epsilon}'\bm{K}_j.
\end{align}
By identical steps, we get 
\begin{align}\label{knockofflikelihood}
    [\bm{y}|\tilde{\delta}_j=1,\bm{\gamma}_{-j}^{(i)}]
    =\left(2\uppi\sigma^2_e\right)^{-n/2}\left(\tilde{\sigma}^2_a\right)^{-1/2}\left(\tilde{a}_{\star}\right)^{1/2}\exp{\left\{-\frac{1}{2\sigma^2_e}\bm{\epsilon}'\bm{\epsilon}+\frac{1}{2 \tilde{a}_{\star}}\tilde{b}_{\star}^2\right\}},
\end{align}
where 
\begin{align}
    \tilde{a}_{\star}&=\left(\frac{\tilde{\bm{K}}_j'\tilde{\bm{K}}_j}{\sigma^2_e}+\frac{1}{\tilde{\sigma}^2_a}\right)^{-1}, \\
    \tilde{b}_{\star}&=\frac{1}{\sigma^2_e}\bm{\epsilon}'\tilde{\bm{K}}_j,
\end{align}
and, we denote $\tilde{\sigma}^2_a$ to emphasize $\tilde{\sigma^2_a}$ in equation (\ref{knockofflikelihood}) is not equal to $\sigma^2_a$ in equation (\ref{originallikelihood}). Lastly, we have 
\begin{align}\label{nulllikelihood}
    [\bm{y}|\nu_j=0,\bm{\gamma}_{-j}^{(i)}]
    =\left(2\uppi\sigma^2_e\right)^{-n/2}\exp{\left\{-\frac{1}{2\sigma^2_e}\bm{\epsilon}'\bm{\epsilon}\right\}}.
\end{align}

Computing the Rao-Blackwellized estimate, (\ref{Rao_Blackwell_Estimate}), for all $n_u$ SNPs is computationally demanding, but note that if we save the MCMC samples of $\bm{\gamma}_{-j}$, we can compute all estimates outside of the MCMC algorithm. To reduce storage requirements, we computed the Rao-Blackwellized estimates within the MCMC algorithm but only for every $n_{\text{thin}}$th iteration, where  $1/n_{\text{thin}}$ was the thinning rate. Because $u_j=0$ when $\delta_j=0$, the individual SNP effects are also prone to high sampling variance, and we obtained samples from their posterior distribution every $n_{\text{thin}}$th iteration,
\begin{align}
    [u_j|\bm{y}]\stackrel{d}{=}\mathcal{N}(a_{\star}b_{\star},a_{\star}).
\end{align}

\baselineskip=10pt
\bibliographystyle{apalike}
\bibliography{bibliography.bib}